\documentclass[preprintnumbers,superscriptaddress,showkeys,showpacs,byrevtex]{revtex4}
\usepackage{amsmath,amsfonts,amssymb,amscd,amsxtra,amsthm}
\usepackage{graphicx}
\usepackage{bm}
\usepackage{amsmath}
\usepackage{multirow}
\newcommand {\be} {\begin{equation}}
\newcommand {\ba} {\begin{eqnarray}}
\newcommand {\ee} {\end{equation}}
\newcommand {\ea} {\end{eqnarray}}
\begin{document}
\preprint{CYCU-HEP-17-09}
\title{Kaon quark distribution functions in the chiral constituent quark model}
\author{Akira Watanabe}
\email[E-mail: ]{akira@ihep.ac.cn}
\affiliation{Institute of High Energy Physics, Chinese Academy of Sciences, Beijing 100049, People's Republic of China}
\affiliation{Theoretical Physics Center for Science Facilities, Chinese Academy of Sciences, Beijing 100049, People's Republic of China}
\affiliation{University of Chinese Academy of Sciences, Beijing 100049, People's Republic of China}
\affiliation{Department of Physics and Center for High Energy Physics, Chung-Yuan Christian University, Chung-Li 32023, Taiwan}
\author{Takahiro Sawada}
\email[E-mail: ]{sawadat@umich.edu}
\affiliation{Department of Physics, University of Michigan, Ann Arbor, Michigan, 48109-1040, USA}
\author{Chung Wen Kao}
\email[E-mail  (Corresponding Author): ]{cwkao@cycu.edu.tw}
\affiliation{Department of Physics and Center for High Energy Physics, Chung-Yuan Christian University, Chung-Li 32023, Taiwan}
\date{\today}
\begin{abstract}
We investigate the valence $u$ and $\bar{s}$ quark distribution functions of the $K^+$ meson,
$v^{K(u)}(x,Q^2)$ and $v^{K(\bar{s})}(x,Q^2)$,
in the framework of the chiral constituent quark model.
We judiciously choose the bare distributions at the initial scale to generate the dressed distributions at the higher scale, considering the meson cloud effects and the QCD evolution, which agree with the phenomenologically satisfactory valence quark distribution of the pion and the experimental data of the ratio $v^{K(u)}(x,Q^2)/v^{\pi (u)}(x,Q^2)$.
We show how the meson cloud effects  affect the bare distribution functions in detail.
We find that a smaller $SU(3)$ flavor symmetry breaking effect is observed, compared with results of the preceding studies based on other approaches.
\end{abstract}
\pacs{12.38.Aw, 13.60.-r, 12.39.-x, 14.40.Aq, 11.10.Hi.}
\keywords{Parton distribution functions, Meson cloud effects, Chiral symmetry breaking, SU(3) flavor symmetry breaking}
\maketitle
\section{Introduction}

The $SU(3)$ flavor symmetry has played an important role in the development of hadron physics since the first strange particle, the kaon, was discovered in 1947.
In particular, various static properties of hadrons made of the light quarks $q=(u,d,s)$ have been successfully explained by virtue of this symmetry.
With the advent of quantum chromodynamics (QCD), we now understand that this success is due to the fact that the masses of the light quarks are small compared with the typical scale of QCD.
However, the $SU(3)$ flavor symmetry is explicitly broken by the strange quark mass which is larger than the up and down quark masses.
One can study this symmetry breaking by considering the pattern of the light quark masses, which provides an important link between QCD and the hadronic structure.
Since the static properties of hadrons are dominated by the long-distance physics which is essentially  nonperturbative in nature, one cannot directly apply the perturbative technique of QCD there, and this fact has attracted a lot of attention for decades.
Besides the static and quasi-static properties of the hadrons such as the polarizabilities and the form factors,
the hadron structure has also been explored in the high energy scattering processes such as the deep inelastic scattering or the Drell-Yan processes.
The partonic structure of hadrons is encoded into the parton distribution functions (PDFs).
Although one cannot calculate them by using QCD itself because of their nonperturbative nature, some indirect ways including employing phenomenological models provide us with opportunities to investigate.
There are several models which have been created and used widely so far.
In this study, we focus on the quark distribution functions in PDFs, and one of the most suitable models to investigate them is the chiral constituent quark model~\cite{Melnitchouk:1994en,Kulagin:1995ia}.\\

In this model, the constituent quarks are surrounded by meson clouds, and one can evaluate the dressing corrections at some low scale $Q_0^2$.
The calculated distribution functions are evolved by the DGLAP equation to higher $Q^2$ at which high energy scattering occurs in practice, and then sea-quarks and gluons emerge as PDFs.
The existence of the meson cloud is based on the fact that the chiral symmetry of the QCD vacuum is spontaneously broken and the light pseudoscalar mesons are the corresponding Nambu-Goldstone bosons.
Those Nambu-Goldstone bosons couple to any particles and develop the meson clouds around the cores.
Therefore the meson cloud effects play dominant roles in the hadronic structures and interactions at low energies.
Since the $SU(3)$ flavor symmetry breaking effect in the Nambu-Goldstone bosons is manifested by the mass difference between the pion and kaon, studying their structures is closely related to the investigation of the symmetry breaking.
It is problematic to apply the leading-order or even the next-to-leading-order (NLO) QCD evolution to the PDFs computed at the initial scale where the value of $\alpha_{s}$ is usually quite large.
Hence it is necessary to take the meson cloud effects into account to understand the hadron structures particularly at such a low energy (see Refs.~\cite{Suzuki:1997wv,Kofler:2017uzq}, and references therein).\\

Focusing on the meson cloud effects, many studies have been done so far to investigate the nucleon PDFs, and there have been various interesting discoveries such as a flavor asymmetry in the PDFs of the nucleon.
This asymmetry is due to the Goldstone boson fluctuations which also generate significant depolarization effects reducing the fraction of the nucleon spin carried by the quarks~\cite{Suzuki:1997wv}.
However, preceding studies on PDFs of the Goldstone bosons are much fewer.
This is because it is rather difficult to extract those PDFs since the pseudoscalar mesons are not stable particles.
Our knowledge of the pion PDFs is entirely from the Drell-Yan di-muon production, $\pi^{\pm}N\rightarrow \mu^{+}\mu^{-}X$~\cite{Conway:1989fs,Bordalo:1987cr,Betev:1985pf}.
Nevertheless, there are still several theoretical studies on the pion PDFs since the investigation of the pion structure is one of the most important subjects in hadron physics because of its Nambu-Goldstone nature~\cite{Aicher:2010cb,Sutton:1991ay,Shigetani:1993dx,Davidson:1994uv,Gluck:1999xe,Weigel:1999pc,Dorokhov:2000gu,Hecht:2000xa, Nguyen:2011jy,Wijesooriya:2005ir,Belitsky:1996vh,Nam:2011hg,Nam:2012af,Nam:2012vm,Chen:2012txa,Chang:2014lva,Chen:2016sno,Alberg:2011yr,Hutauruk:2016sug}, and some related lattice studies have also been done~\cite{Daniel:1990ah,Detmold:2003tm}.
In particular, the authors of Ref.~\cite{Aicher:2010cb} performed their analysis including the next-to-leading-logarithmic threshold resummation effects in the calculation of the Drell-Yan cross section to extract the valence quark distribution of the pion.
We have found that their results can be reproduced within the chiral constituent quark model, and investigated the roles of the meson cloud effects in the pion valence quark distribution~\cite{Watanabe:2016lto}.
Hence, it is natural for us to extend our study to the kaon case to see the meson cloud effects and the $SU(3)$ flavor symmetry breaking effect in the kaon quark distribution functions.
The experimental extraction of the kaon PDFs is even more difficult, and currently only the data for the ratio of the valence quark distributions of the pion and kaon presented in Ref.~\cite{Badier:1980jq} are available.
However, the kaon-induced Drell-Yan experiment has been proposed at some experimental facilities.
Our predictions presented in this article can be tested there in the future.\\

In this study, we assume the functional forms of the bare quark distributions of the kaon with some adjustable parameters at the initial scale, which is the so-called model scale.
We evaluate the dressing corrections to the bare distributions at this scale, and obtain the dressed distributions at higher $Q^2$ by performing the QCD evolution.
The parameters are fixed to reproduce the experimental data of the ratio of the pion and kaon valence quark distributions.
As to the required bare distribution of the pion, we utilize the results of the previous work~\cite{Watanabe:2016lto}.
The resulting $u$ and $\bar s$ quark distributions of the $K^+$ meson will be shown, and then we will discuss in detail the features of the meson cloud effects and the $SU(3)$ flavor symmetry breaking effect in those distributions.\\

This article is organized as follows:
In Sec.~II, we briefly explain how to evaluate the dressing corrections in the chiral constituent quark model, and present the resulting analytical expressions of the dressed distributions.
We show and discuss our numerical results in Sec.~III.
Sec.~IV is devoted to the conclusion.\\

\section{Dressing corrections to quark distribution functions of the kaon}
\begin{figure}[t]
\begin{tabular}{c}
\includegraphics[width=0.8\textwidth]{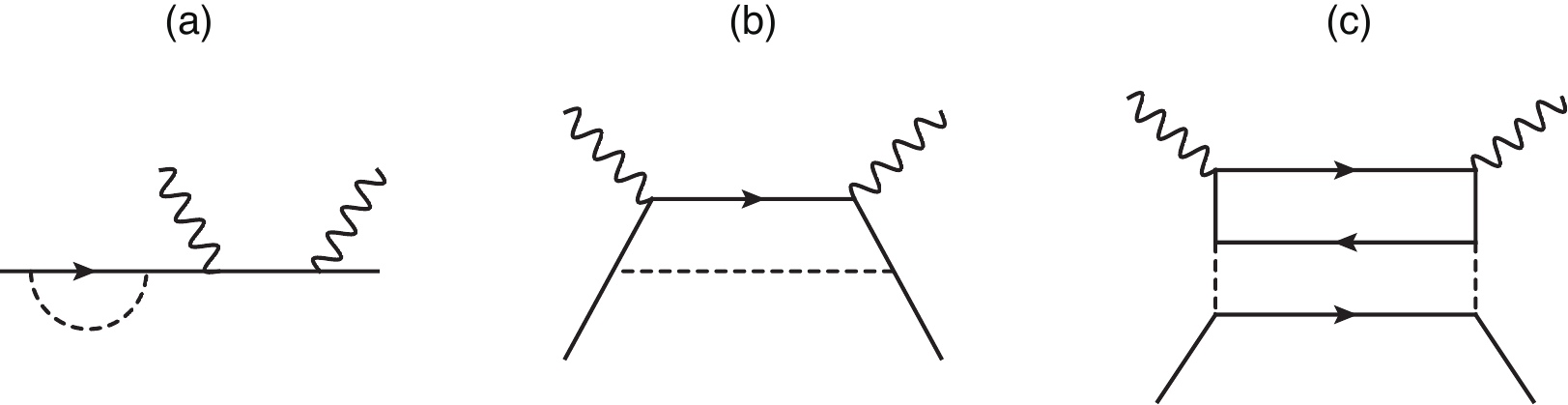}
\end{tabular}
\caption{
The three diagrams characterizing the dressing corrections to a constituent quark.
The wavy, thick, and dashed lines denote the virtual photons, the constituent quarks, and the Goldstone bosons, respectively.
}
\label{fig:diagrams}
\end{figure}

In this section, we briefly sketch our scheme of computing the dressing corrections from the meson cloud effects on constituent quarks in the framework of the chiral constituent quark model.
Its applications to the nucleon and the pion were demonstrated in Ref.~\cite{Suzuki:1997wv} and Ref.~\cite{Watanabe:2016lto}, respectively.
The constituent quark field $\psi$ and the Nambu-Goldstone boson fields $\Pi$ are defined as
\be
\psi=\begin{pmatrix}
u\\
d\\
s\\
\end{pmatrix}
,\,\,\,\,
\Pi=\frac{1}{\sqrt{2}}\begin{pmatrix}
\frac{\pi^{0}}{\sqrt{2}}+\frac{\eta}{\sqrt{6}} & \pi^{+} & K^{+} \\
\pi^{-} & -\frac{\pi^{0}}{\sqrt{2}}+\frac{\eta}{\sqrt{6}} & K^{0} \\
K^{-} &\bar{K}^{0} & -\frac{2\eta}{\sqrt{6}}\\
\end{pmatrix}.
\ee
Their interactions are described by a simple effective Lagrangian with the full $SU(3)$ flavor symmetry,
\be
{\mathcal{L}}_{int}=-\frac{g_A}{f}\bar{\psi}\gamma^{\mu}\gamma_{5}(\partial_{\mu}\Pi)\psi,
\ee
where $f$ and $g_A$ represent the pseudoscalar decay constant and the quark axial-vector coupling constant, respectively.
Assuming that the isospin symmetry is exact, we have the following relations between
the dressed and bare states of constituent quarks,
\ba
|U\rangle &=& \sqrt{Z_u}|u_0\rangle +a_{\pi}|d\pi^{+}\rangle +\frac{a_{\pi}}{2}|u\pi^{0}\rangle +a_K|sK^{+}\rangle +\frac{a_\eta}{6}|u\eta\rangle,\nonumber \\
|D\rangle &=& \sqrt{Z_d}|d_0\rangle +b_{\pi}|u\pi^{-}\rangle +\frac{b_{\pi}}{2}|d\pi^{0}\rangle +b_K|sK^{0}\rangle +\frac{b_\eta}{6}|d\eta\rangle , \nonumber \\
|S\rangle &=& \sqrt{Z_{s}}|s_0\rangle +c_{K}|u K^{-}\rangle +c_{K}|d \bar{K}^{0}\rangle  +\frac{2c_\eta}{3}|s\eta\rangle ,
\label{eq:dressed_states}
\ea
where $Z_u = Z_d$ and $Z_s$ are the renormalization constants for the bare constituent quarks.
$|a_{\alpha}|^2 = |b_{\alpha}|^2$ and $|c_{\alpha}|^2$ represent the probabilities of finding a Goldstone boson $\alpha$ in the dressed states of constituent $u$, $d$, and $s$ quarks, respectively.
The dressing corrections are characterized by the three diagrams in Fig.~\ref{fig:diagrams}.
The contributions of the renormalization constants are described by Fig.~\ref{fig:diagrams}~(a).
Fig.~\ref{fig:diagrams}~(b) depicts the process in which the virtual photons directly couple to the constituent quark and the Goldstone boson around the constituent quark is just a spectator.
Fig.~\ref{fig:diagrams}~(c) shows the process in which the virtual photons couple to the sea-quark pair emitted by the Goldstone boson dressing the constituent quark.
In this work, we evaluate those diagrams in the infinite momentum frame, because the $Z$-graph terms vanish in this frame and the calculations are simplified.\\

The dressing corrections can be analytically expressed by the convolution integrals, and through this article we adopt the following short hand notation,
\be
P \otimes q \equiv \int_x^1 {\frac{{dy}}{y}} P\left( y \right)q\left( {\frac{x}{y}} \right).
\ee
The contributions corresponding to Fig.~\ref{fig:diagrams}~(b) and (c) are given by ${P_{j\alpha /i}}\otimes {q}_{i}$ and $V_{k /\alpha}\otimes {P_{\alpha j/i}}\otimes {q}_{i}$, respectively.
Here $V_{k/\alpha }$ is the quark distribution function of the Goldstone boson $\alpha$ with the normalization condition,
\be
\int_0^1 {dx} {V_{k/\alpha }}\left( x \right) = 1.
\ee
$P_{j\alpha /i}(y)$ is the splitting function which gives the probability of finding a constituent quark $j$ having the momentum fraction $y$ together with a spectator Goldstone boson $\alpha$ in a parent constituent quark $i$.
Its functional form is given by
\be
{P_{j\alpha /i}}\left( y \right)  =  \frac{1}{{8{\pi ^2}}}{\left( {\frac{{{g_A}\bar m}}{f}} \right)^2}\int {dk_T^2} \frac{{{{\left( {{m_j} - {m_i}y} \right)}^2} + k_T^2}}{{{y^2}\left( {1  -  y} \right){{\left( {m_i^2 - M_{j\alpha }^2} \right)}^2}}},
\label{eq:Pjalphaoveri}
\ee
where $P_{\alpha j/i} (x) = P_{j \alpha /i} (1 - x)$, and $m_i$, $m_j$, and $m_{\alpha}$ represent the masses of the constituent quarks and the spectator Goldstone boson.
$M_{j\alpha }^2$ is the invariant mass squared of the final state,
\be
M_{j\alpha }^2  =  \frac{{m_j^2 + k_T^2}}{y} + \frac{{m_\alpha ^2 + k_T^2}}{{1  -  y}},
\ee
and ${\bar m} = (m_i + m_j )/2$ represents the average of the constituent quark masses.
Since the integral in Eq.~\eqref{eq:Pjalphaoveri} diverges if performed straightforwardly, we need to replace $g_A$ with the following function to make the integral to converge~\cite{Speth:1996pz},
\be
{g_A}\exp \left( {\frac{{m_i^2 - M_{j\alpha }^2}}{{4{\Lambda ^2}}}} \right),
\ee
where $\Lambda$ is the cutoff.
Defining the moments of the splitting functions as
\be
\left\langle x^{n-1} P_{j\alpha /i} \right\rangle = \left\langle {{x^{n-1} P_{\alpha j/i}}} \right\rangle \equiv \int_0^1 {dx} {x^{n-1} P_{j\alpha /i}}\left( x \right) ,
\ee
we express the first moments as $\left\langle P_\alpha \right\rangle$ hereafter.
The values of the adjustable parameters that we choose for the numerical evaluations in this work are shown in Table~\ref{table:parameters}.
\begin{table}[bt]
\caption{The set of parameters chosen in this work.}
\begin{center}
\begin{tabular}{| c | c | c | c | c | c | c | c | c |} \hline
$g_A$ & $m_u$ & $m_d$ & $m_s$ & $m_\pi$ & $m_K$ & $m_\eta$ & $f_\pi$ & $\Lambda$ \\
\hline \hline
1.0 & 360 MeV & 360 MeV & 570 MeV & 140 MeV & 494 MeV & 548 MeV & 93 MeV & 1.4 GeV \\
\hline
\end{tabular}
\end{center}
\label{table:parameters}
\end{table}
They are typical values guided by the related preceding studies based on the Nambu--Jona-Lasinio model~\cite{Nambu:1961tp, Nambu:1961fr, Hatsuda:1994pi}.
In particular, the obtained value of the Gottfried sum rule by using this parameter set in the present scheme agrees with the empirical value~\cite{Suzuki:1997wv}.\\

Here we present the obtained expressions of the dressed valence $u$ and $\bar s$ quark distributions of the $K ^+$ meson.
Their bare quark distribution functions, $u_0 (x)$ and $\bar s _0 (x)$, at the initial scale $Q^2 = Q_0^2$ satisfy the normalization conditions,
\be
\int_0^1 {dx} {u_0}\left( x \right) = \int_0^1 {dx} {\bar{s}_0}\left( x \right) = 1.
\nonumber
\ee
First, we show the results for the $u$ quark.
Collecting all the corresponding terms described by the diagrams in Fig.~\ref{fig:diagrams}, the dressed $u$ and $\bar u$ quark distribution functions can be written as
\begin{align}
u\left( x \right) = &Z_{u}{u_0}\left( x \right) + \frac{1}{2}{P_{u\pi /u}} \otimes {u_0} + {V_{u/\pi }} \otimes {P_{\pi d/u}} \otimes {u_0} + \frac{1}{4}{V_{u/\pi }} \otimes {P_{\pi u/u}} \otimes {u_0} \nonumber \\
&+ {V_{u/K}} \otimes {P_{Ks/u}} \otimes {u_0} + {V_{u/K}} \otimes {P_{K\bar u/\bar s}} \otimes {{\bar s}_0} \nonumber \\
&+ \frac{1}{6}{P_{u\eta /u}} \otimes {u_0} + \frac{1}{{36}}{V_{u/\eta }} \otimes {P_{\eta u/u}} \otimes {u_0} + \frac{1}{9}{V_{u/\eta }} \otimes {P_{\eta \bar s/\bar s}} \otimes {{\bar s}_0},\nonumber  \\
\bar u\left( x \right) = &{P_{\bar uK/\bar s}} \otimes {{\bar s}_0} + \frac{1}{4}{V_{\bar u/\pi }} \otimes {P_{\pi u/u}} \otimes {u_0} \nonumber \\
&+ \frac{1}{{36}}{V_{\bar u/\eta }} \otimes {P_{\eta u/u}} \otimes {u_0} + \frac{1}{9}{V_{\bar u/\eta }} \otimes {P_{\eta \bar s/\bar s}} \otimes {{\bar s}_0},
\end{align}
respectively.
The renormalization constant is expressed as
\be
Z_u = 1 - \frac{3}{2}\left\langle {{P_\pi }} \right\rangle  - \left\langle {{P_{K (i=u)}}} \right\rangle - \frac{1}{6} \left\langle {{P_{\eta (i=u)}}} \right\rangle ,
\label{eq:renormalization_constant_for_u}
\ee
where $\left\langle {{P_{K (i=u)}}} \right\rangle$ and $\left\langle {{P_{\eta (i=u)}}} \right\rangle$ denote the first moments of the splitting functions in which the parent constituent quark is a $u$ quark.
Hence, we obtain the expression of the dressed valence $u$ quark distribution function,
\begin{align}
v^{K(u)}_{\rm dressed} \left( x \right) = &u\left( x \right) - \bar u\left( x \right) \nonumber \\
= &Z_u {u_0}\left( x \right) + \frac{1}{2}{P_{u\pi /u}} \otimes {u_0} + {V_{u/\pi }} \otimes {P_{\pi d/u}} \otimes {u_0} + {V_{u/K}} \otimes {P_{Ks/u}} \otimes {u_0} \nonumber \\
&+ {V_{u/K}} \otimes {P_{K\bar u/\bar s}} \otimes {{\bar s}_0} - {P_{\bar uK/\bar s}} \otimes {{\bar s}_0} + \frac{1}{6}{P_{u\eta /u}} \otimes {u_0}. \label{eq:uval_in_kplus}
\end{align}
Integrating both sides of Eq.~\eqref{eq:uval_in_kplus}, we obtain
\begin{align}
\int_0^1 {dx} v^{K(u)}_{\rm dressed} \left( x \right)
= &Z_u + \frac{3}{2}\left\langle {{P_\pi }} \right\rangle  + \left\langle {{P_{K\left( {i = u} \right)}}} \right\rangle  + \frac{1}{6}\left\langle {{P_{\eta \left( {i = u} \right)}}} \right\rangle \nonumber \\
= &1,
\nonumber
\end{align}
which shows the correct normalization.\\

To investigate the meson cloud effects on the valence $u$ quark distribution in detail, we decompose the dressed distribution into several terms as follows:
\be
{v^{K(u)}_{\rm dressed}}\left( x \right) = {v^{K(u),a}}\left( x \right) + {v^{K(u),b1}}\left( x \right) + {v^{K(u),b2}}\left( x \right) + {v^{K(u),b3}}\left( x \right) + {v^{K(u),c1}}\left( x \right) + {v^{K(u),c2}}\left( x \right) ,
\ee
where
\begin{align}
{v^{K(u),a}}\left( x \right) &= {Z_u}{u_0}\left( x \right), \nonumber \\
{v^{K(u),b1}}\left( x \right) &= \frac{1}{2}{P_{u\pi /u}} \otimes {u_0},\nonumber \\
{v^{K(u),b2}}\left( x \right) &=  - {P_{\bar uK/\bar s}} \otimes {{\bar s}_0},\nonumber \\
{v^{K(u),b3}}\left( x \right) &= \frac{1}{6}{P_{u\eta /u}} \otimes {u_0}, \nonumber \\
{v^{K(u),c1}}\left( x \right) &= {V_{u/\pi }} \otimes {P_{\pi d/u}} \otimes {u_0}, \nonumber \\
{v^{K(u),c2}}\left( x \right) &= {V_{u/K}} \otimes {P_{Ks/u}} \otimes {u_0} + {V_{u/K}} \otimes {P_{K\bar u/\bar s}} \otimes {{\bar s}_0} .
\label{eq:eachterms_u}
\end{align}
Here the first term ${v^{K(u),a}}\left( x \right)$ represents the contribution from the renormalization constant only which corresponds to Fig.~\ref{fig:diagrams}~(a), and the upper indices $b$ and $c$ correspond to the diagrams in Fig.~\ref{fig:diagrams}~(b) and (c), respectively.
The numbers 1, 2, and 3 denoted as the upper indices correspond to the contributions in which the involved Goldstone bosons are the pion, the kaon, and the $\eta$ meson, respectively.
We display and discuss their contributions individually in the next section.\\

Next, we present the analytical results for the $\bar s$ quark in the $K^+$ meson.
The dressed distribution functions of the $\bar s$ and $s$ quarks can be written as
\begin{align}
\bar s\left( x \right) = &{Z_{s}}{{\bar s}_0}\left( x \right) + {V_{\bar s/K}} \otimes {P_{Ks/u}} \otimes {u_0} + {V_{\bar s/K}} \otimes {P_{K\bar u/\bar s}} \otimes {{\bar s}_0} + {V_{\bar s/K}} \otimes {P_{K\bar d/\bar s}} \otimes {{\bar s}_0} \nonumber \\
&+ \frac{2}{3}{P_{\bar s\eta /\bar s}} \otimes {{\bar s}_0} + \frac{1}{9}{V_{\bar s/\eta }} \otimes {P_{\eta u/u}} \otimes {u_0} + \frac{4}{9}{V_{\bar s/\eta }} \otimes {P_{\eta \bar s/\bar s}} \otimes {{\bar s}_0}, \nonumber \\
s\left( x \right) = &{P_{sK/u}} \otimes {u_0} + \frac{1}{9}{V_{s/\eta }} \otimes {P_{\eta u/u}} \otimes {u_0} + \frac{4}{9}{V_{s/\eta }} \otimes {P_{\eta \bar s/\bar s}} \otimes {{\bar s}_0},
\end{align}
respectively.
The renormalization constant is expressed as
\be
Z_{s} = 1 - 2 \left\langle {{P_{K (i=\bar{s})}}} \right\rangle - \frac{2}{3} \left\langle {{P_{\eta (i=\bar s )}}} \right\rangle . \label{eq:renormalization_constant_for_sbar}
\ee
Hence, we obtain the expression of the dressed valence $\bar{s}$ quark distribution function,
\begin{align}
v^{K(\bar{s})}_{\rm dressed} \left( x \right) = &\bar{s}\left( x \right) - s\left( x \right) \nonumber \\
= &{Z_{s}}{{\bar s}_0}\left( x \right) + {V_{\bar s/K}} \otimes {P_{Ks/u}} \otimes {u_0} + {V_{\bar s/K}} \otimes {P_{K\bar u/\bar s}} \otimes {{\bar s}_0} + {V_{\bar s/K}} \otimes {P_{K\bar d/\bar s}} \otimes {{\bar s}_0} \nonumber \\
&+ \frac{2}{3}{P_{\bar s\eta /\bar s}} \otimes \bar s - {P_{sK/u}} \otimes {u_0}. \label{eq:sbarval_in_kplus}
\end{align}
Integrating both sides of Eq.~\eqref{eq:sbarval_in_kplus}, we obtain
\begin{align}
\int_0^1 {dx} v^{K(\bar{s})}_{\rm dressed} \left( x \right)
= &{Z_{s}} + 2\left\langle {{P_{K\left( {i = \bar s} \right)}}} \right\rangle  + \frac{2}{3}\left\langle {{P_{\eta \left( {i = \bar s} \right)}}} \right\rangle \nonumber \\
= &1,
\nonumber
\end{align}
which shows the correct normalization.\\

Similar to the $u$ quark case, we decompose the dressed valence $\bar s$ quark distribution into several terms as follows:
\be
{v^{K(\bar s)}}\left( x \right) = {v^{K(\bar s),a}}\left( x \right) + {v^{K(\bar s),b2}}\left( x \right) + {v^{K(\bar s),b3}}\left( x \right) + {v^{K(\bar s),c2}}\left( x \right),
\ee
where
\begin{align}
{v^{K(\bar s),a}}\left( x \right) &= {Z_{s}}{{\bar s}_0}\left( x \right), \nonumber \\
{v^{K(\bar s),b2}}\left( x \right) &=  - {P_{sK/u}} \otimes {u_0}, \nonumber \\
{v^{K(\bar s),b3}}\left( x \right) &= \frac{2}{3}{P_{\bar s\eta /\bar s}} \otimes {{\bar s}_0},\nonumber \\
{v^{K(\bar s),c2}}\left( x \right) &= {V_{\bar s/K}}{P_{Ks/u}} \otimes {u_0} + {V_{\bar s/K}} \otimes {P_{K\bar u/\bar s}} \otimes {{\bar s}_0} + {V_{\bar s/K}} \otimes {P_{K\bar d/\bar s}} \otimes {{\bar s}_0}.
\label{eq:eachterms_s}
\end{align}
The notations are the same as those in the $u$ quark case, and the discussion on the individual terms is given in the next section.\\

Here we make some remarks on the values of  $\langle P_\alpha \rangle $ and the renormalization constants, which are obtained with the parameter set shown in Table~\ref{table:parameters} and presented in Table~\ref{table2}.
\begin{table}[bt]
\caption{The resulting values of the first moments of the splitting functions and the renormalization constants.}
\begin{center}
\begin{tabular}{| c | c | c | c | c | c | c |} \hline
$\langle P_{\pi}\rangle  $ & $\langle P_{K(i=u)}\rangle $ &  $\langle P_{K(i=\bar{s})}\rangle $ & $\langle P_{\eta(i=u)}\rangle$  & $\langle P_{\eta(i=\bar{s})}\rangle $
& $Z_{u}$ & $Z_{s}$ \\
\hline
0.16 & 0.09 & 0.14 & 0.06 & 0.12 &0.67 &0.63 \\
\hline
\end{tabular}
\end{center}
\label{table2}
\end{table}
We can clearly observe the $SU(3)$ flavor symmetry breaking in the values of $\langle P_{\alpha}\rangle $, but the difference between $Z_u$ and $Z_{s}$ is relatively small.
It should be noted that the $\eta$ meson contributions to $Z_u$ and $Z_{s}$ are quite different from each other, and this difference reduces the difference between the two renormalization constants.
Considering the factors in the right-hand sides of Eqs.~\eqref{eq:renormalization_constant_for_u} and \eqref{eq:renormalization_constant_for_sbar}, the $\eta$ meson contribution to $Z_u$ is about 0.01 only, but that to $Z_{s}$ is about 0.08.
This fact implies that the $\eta$ meson contribution should not be neglected when we consider the strange quark distribution function of the kaon, while it was neglected in Ref.~\cite{Watanabe:2016lto} because only the pion structure was studied there.\\

\section{Numerical results and discussion}
In this section, we present our numerical results and discuss their implications.
To obtain the dressed quark distributions of the kaon, we need to determine the bare distribution functions of both the pion and kaon.
As to the pion's, we simply utilize the result of the previous work~\cite{Watanabe:2016lto},
\be
v^{\pi}_{\rm bare}(x,Q_0^2) = N_{\pi}x^{\alpha}(1-x)^{\alpha},
\label{eq:bare_for_pion}
\ee
where $ N_{\pi}$ is the normalization factor and $\alpha = 1.8$.
The value of $\alpha$ was fixed to reproduce the results of Ref.~\cite{Aicher:2010cb} at $Q^2 = 16$~GeV$^2$.
Also, in the previous work the initial scale was fixed as $Q^2_0 = 0.25$~GeV$^2$, and we adopt the same value in this study.\\

Therefore, our task at this stage is to determine the bare valence $u$ and $\bar s$ quark distribution functions of the $K^+$ meson at the initial scale.
To do so, we assume that they own the following functional forms,
\begin{align}
v^{K(u)}_{\rm bare} (x,Q_0^2) &= N_{K(u)} x^{\beta} (1 - x)^{\gamma}, \nonumber \\
v^{K(\bar{s})}_{\rm bare}(x,Q_0^2) &= N_{K(\bar{s})} x^{\gamma} (1 - x)^{\beta},
\label{eq:bare}
\end{align}
respectively.
Here $\beta$ and $\gamma$ are the parameters to be determined, and $N_{K(u)}$ and $N_{K(\bar{s})}$ are the normalization factors.
In this study, to determine those parameters we utilize the experimental data for the ratio of the valence quark distribution functions of the pion and kaon,
$v^{K(u)} (x,Q^2) / v^{\pi(u)} (x,Q^2)$, at $Q^2 = 27$~GeV$^2$~\cite{Badier:1980jq}.
First, we evaluate the dressing corrections to the bare $u$ quark distribution function of the kaon with arbitrarily fixed parameters at the initial scale, and then perform the QCD evolution to obtain the dressed distribution at $Q^2 = 27$~GeV$^2$.
Involving the dressed distribution of the pion, we compare the obtained ratio with that from the data at this scale.
For this procedure, we utilize the DGLAP evolution code~\cite{Kobayashi:1994hy}, and the results presented here are obtained at NLO accuracy.
For the fitting, the MINUIT package~\cite{James:1975dr} is used.
The best fit values of the parameters are found to be:
\begin{align}
\beta &= 1.84\pm0.27, \nonumber \\
\gamma &= 2.01\pm0.19.
\label{eq:fit_result}
\end{align}\\

\begin{figure}[tb]
\begin{center}
\includegraphics[width=0.55\textwidth]{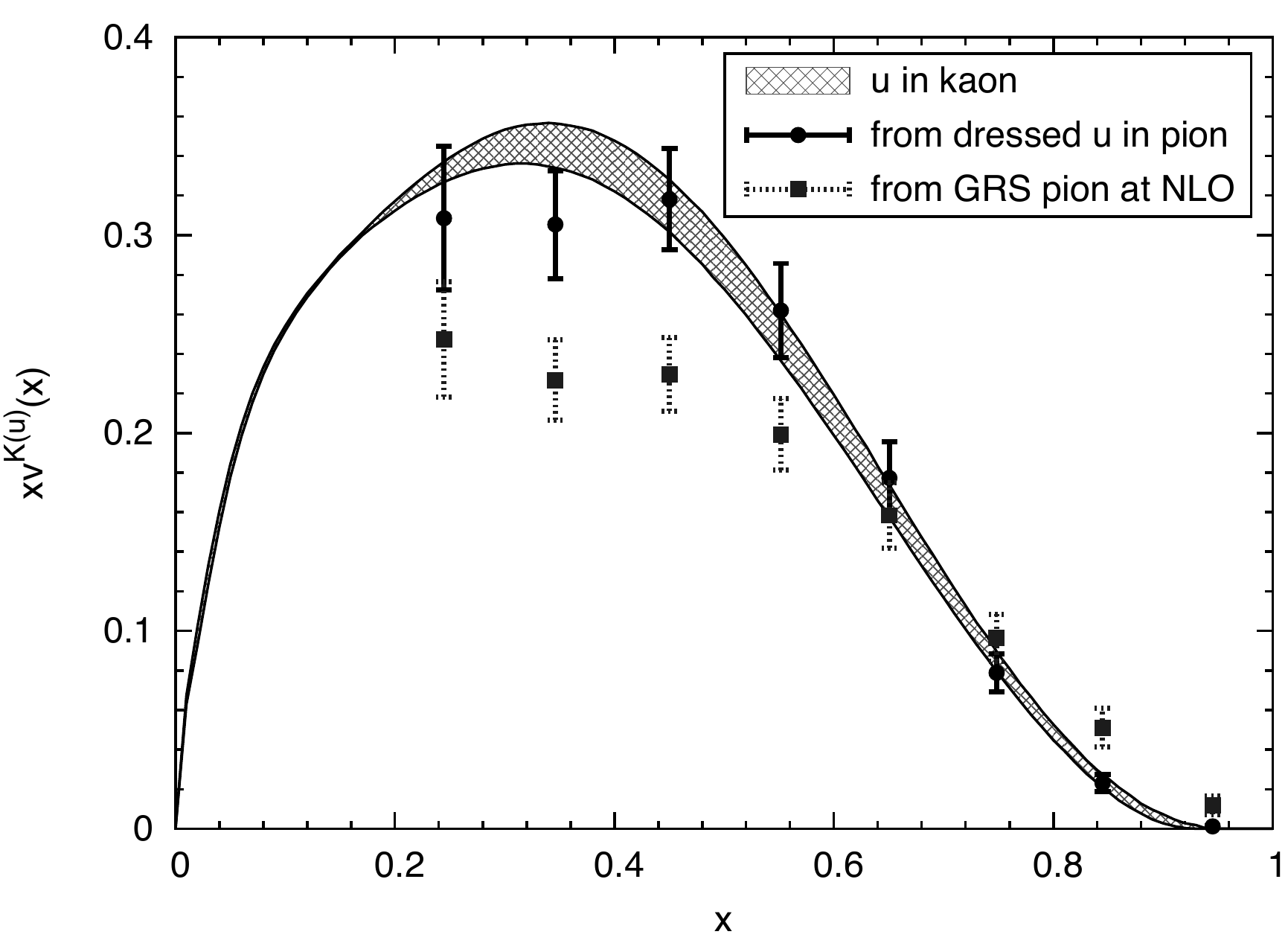}
\caption{
The cross-hatched area represents the 68\% C.L. uncertainty of our dressed valence $u$ quark distribution function of the kaon at $Q^2 = 27$~GeV$^2$.
The circles are the empirical values of  $xv^{K(u)}(x,Q^2)$ from the combination of the experimental data of the ratio $v^{K(u)} / v^{\pi(u)}$~\cite{Badier:1980jq} and the valence quark distribution function of the pion taken from Ref.~\cite{Watanabe:2016lto}.
The squares represent the results obtained from the combination of the ratio and the GRS parametrization~\cite{Gluck:1999xe}. The error bars for those empirical values are also provided here.
}
\label{fig:fit}
\end{center}
\end{figure}

We show in Fig.~\ref{fig:fit}
the resulting valence $u$ quark distribution function of the $K^+$ meson at $Q^2 = 27$~GeV$^2$ by the cross-hatched pattern.
The results depicted by the circles and squares are obtained by using the valence quark distributions of the pion multiplied by the ratio $v^{K(u)} / v^{\pi(u)}$.
For the former ones, we utilize the pion's distribution presented in Ref.~\cite{Watanabe:2016lto}, and the Gl\"uck-Reya-Schienbein (GRS) parametrization~\cite{Gluck:1999xe} is used for the latter ones.
One can find from this figure that the empirical values depicted by the circles are almost perfectly consistent with our results denoted by the cross-hatched pattern within their errors, which means that our fitting is successfully performed.
It should be emphasized here that there are only two adjustable parameters, $\beta$ and $\gamma$, in this fitting, although basically more parameters appears in other models.
It is clear to see that the values based on the GRS parametrization are smaller than the ones from Ref.~\cite{Watanabe:2016lto} when $x\le 0.7$ but become larger when $x\ge 0.7$.
In particular, we notice that the value of the valence quark distribution function of the pion is almost zero at $x=0.93$ in Ref.~\cite{Watanabe:2016lto}, so the $u$ quark distribution function of the kaon also vanishes at this value of $x$. \\

\begin{figure}[tb]
\begin{tabular}{cc}
\includegraphics[width=0.5\textwidth]{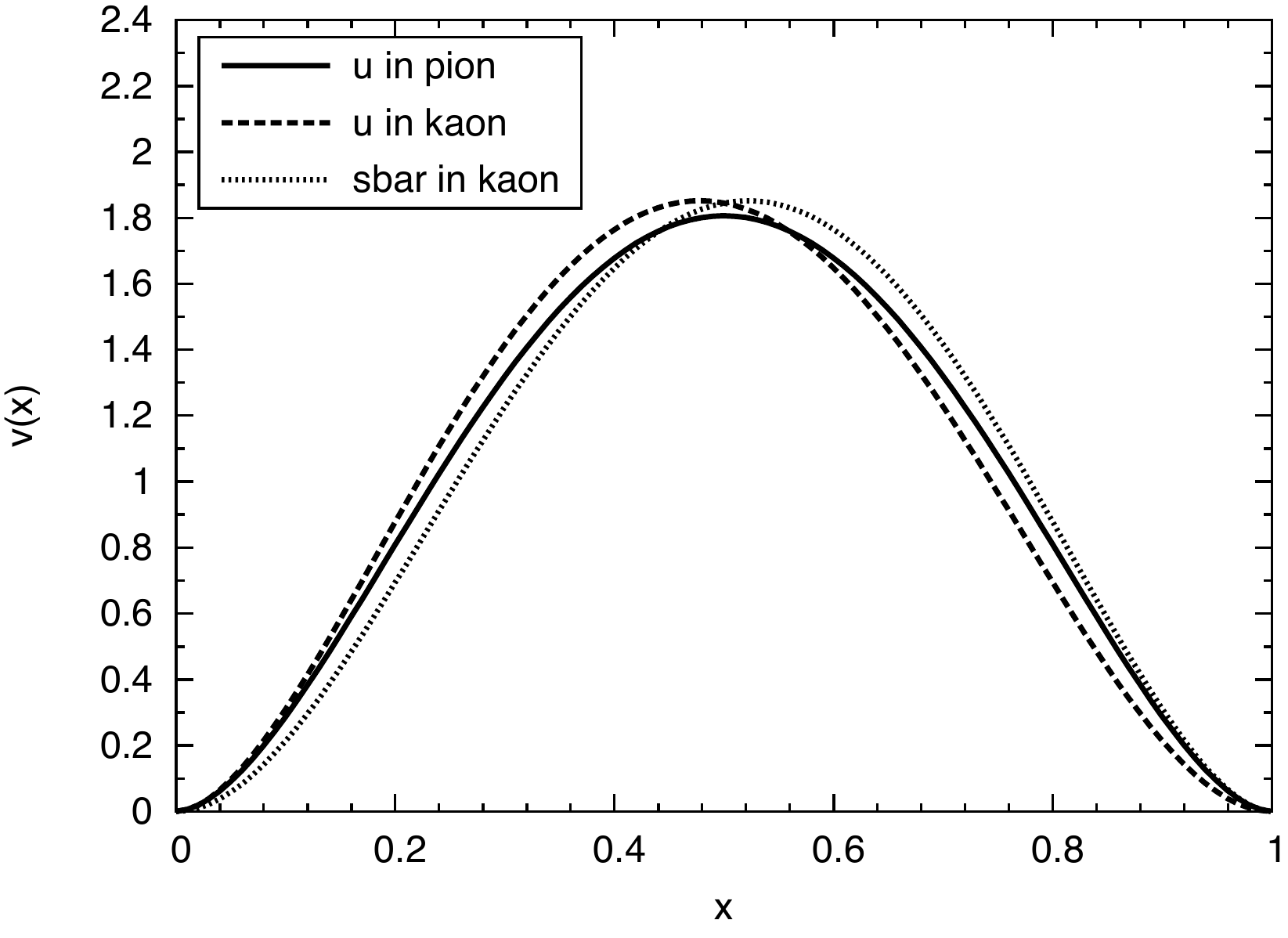}
\includegraphics[width=0.5\textwidth]{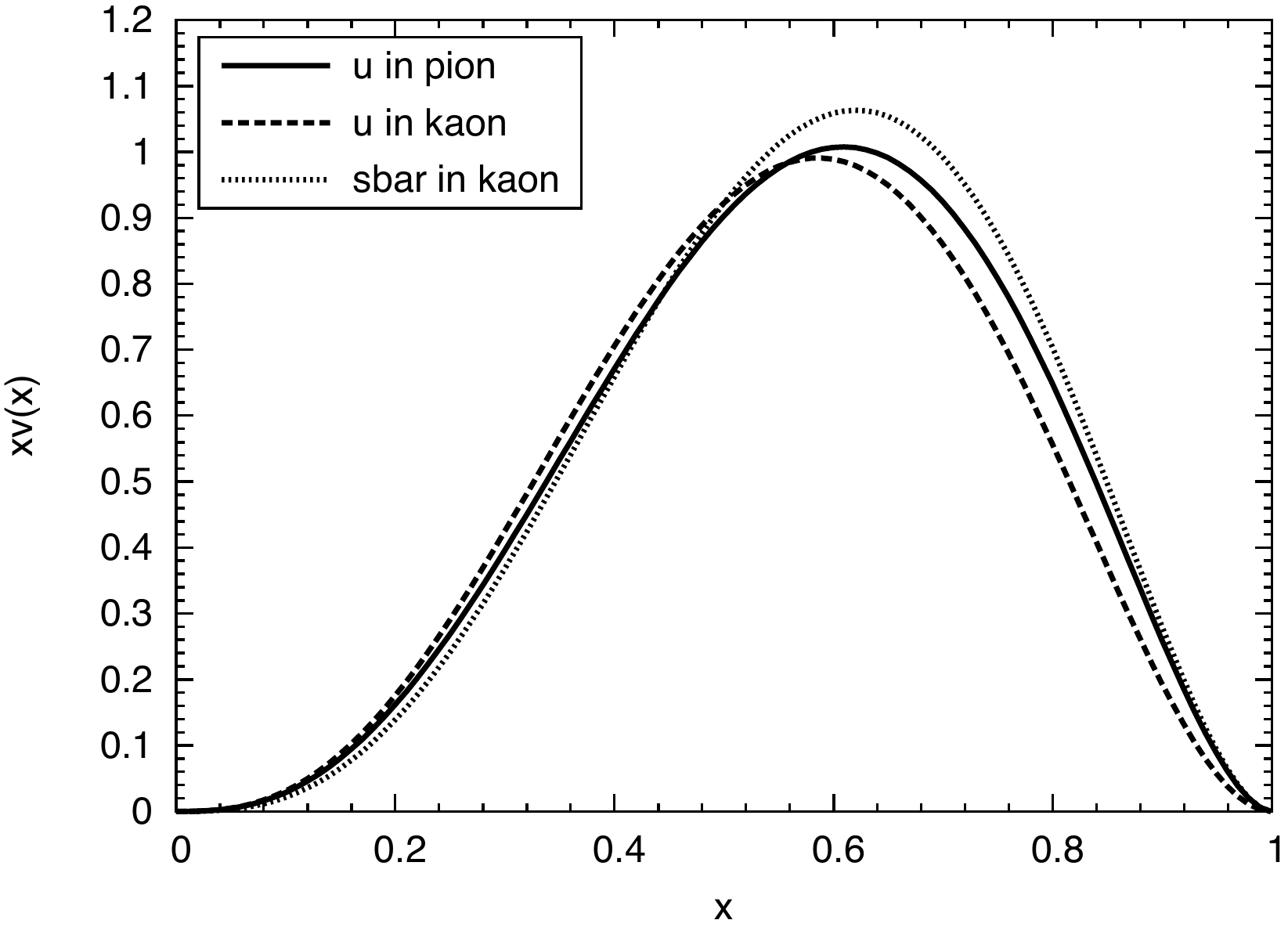}
\end{tabular}
\caption{
The bare valence quark distribution functions of the pion and kaon at $Q^2=Q_0^2$.
The solid lines depict the pion's distribution, and the dashed and dotted lines denote the $u$ and $\bar{s}$ valence quark distributions of the kaon, respectively.
The left panel is for $v^{\pi,K}$ and the right one is for $xv^{\pi,K}$.
}
\label{fig:bareVQDF}
\end{figure}

Using the Eq.~\eqref{eq:sbarval_in_kplus}, with the same values of $\beta$ and $\gamma$ we also can proceed to calculate the dressed valence $\bar{s}$ quark distribution function of the $K^+$ meson.
Although there is no available experimental data of $v^{K(\bar{s})}$ so far, we expect there will be some available data in the near future.
Our resulting $v^{K(u,\bar{s})}_{\rm bare} (x,Q_0^2)$ (left panel) and  $x v^{K(u,\bar{s})}_{\rm bare} (x,Q_0^2)$ (right panel) are displayed in Fig.~\ref{fig:bareVQDF}.
%
%
We also put the bare $u$ quark distribution functions in the figure for comparison.
The peak positions of $v^{K(u)}_{\rm bare}$ and $v^{K(\bar{s})}_{\rm bare}$ are located around $x=0.48$
and $0.52$,
respectively, not exactly at the center.
$x v^{K(u)}_{\rm bare}$ is slightly larger than $x v^{K(\bar{s})}_{\rm bare}$ when $x\le 0.5$, but when $x\ge 0.5$ the situation is substantially inverted.
This observation is qualitatively consistent with those can be seen in the preceding studies~\cite{Nam:2012vm,Chen:2016sno}.
However, we also find that the observed $SU(3)$ flavor symmetry breaking in our bare distributions is much smaller than those found in them.
In particular, we notice that our $v^{\pi}_{\rm bare}$ is quite close to $v^{K(\bar{s})}_{\rm bare}$ in the large $x$ region.
It can be understood as the consequence of the fact that our values of $\alpha$ and $\beta$ in Eqs.~\eqref{eq:bare_for_pion} and \eqref{eq:bare} are close to each other.
The most important task of this work is to figure out how the meson cloud effects affect the $SU(3)$ flavor symmetry breaking in the valence quark distribution functions of the kaon. \\

\begin{figure}[tb]
\begin{tabular}{cc}
\includegraphics[width=0.5\textwidth]{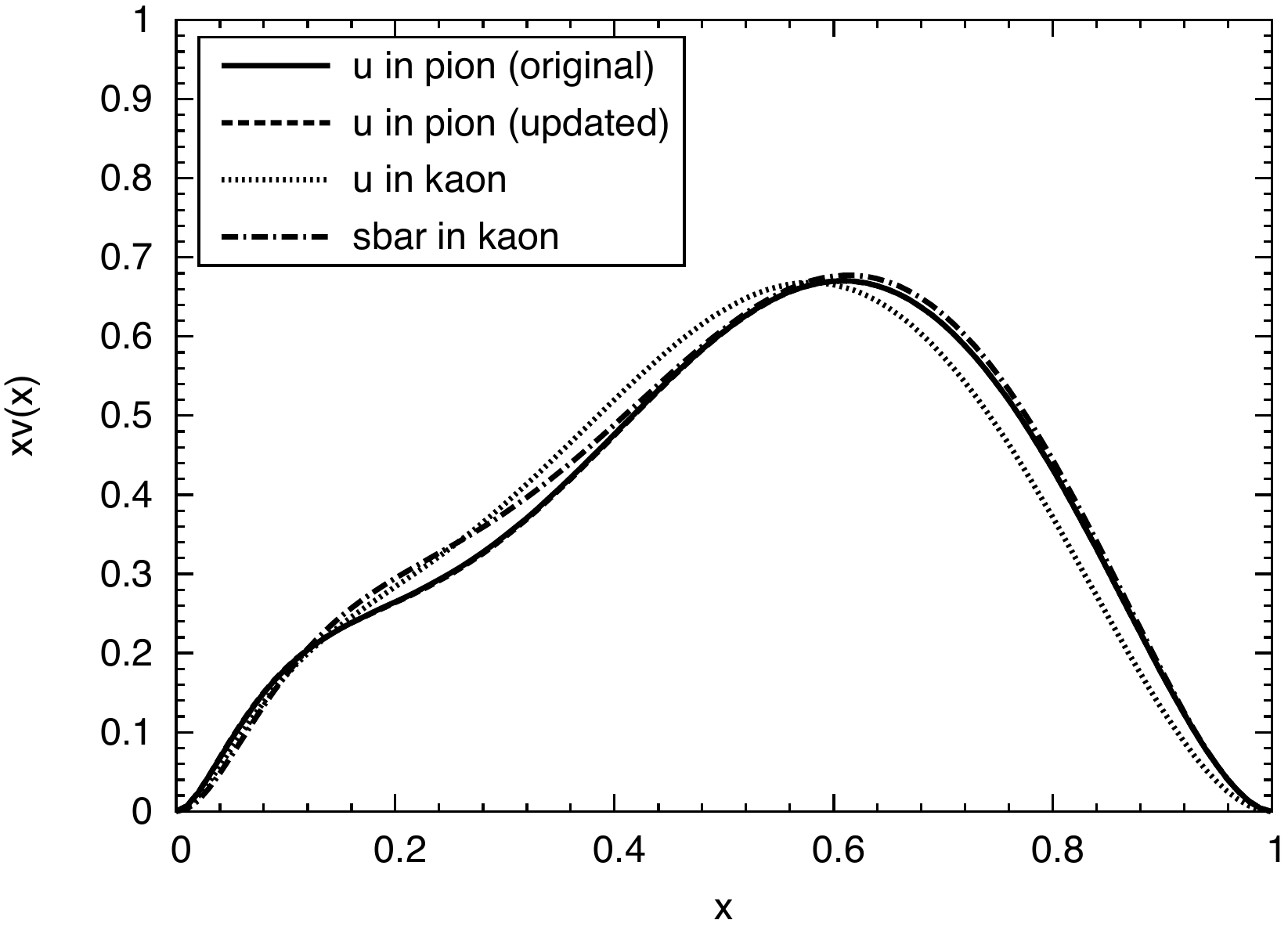}
\includegraphics[width=0.5\textwidth]{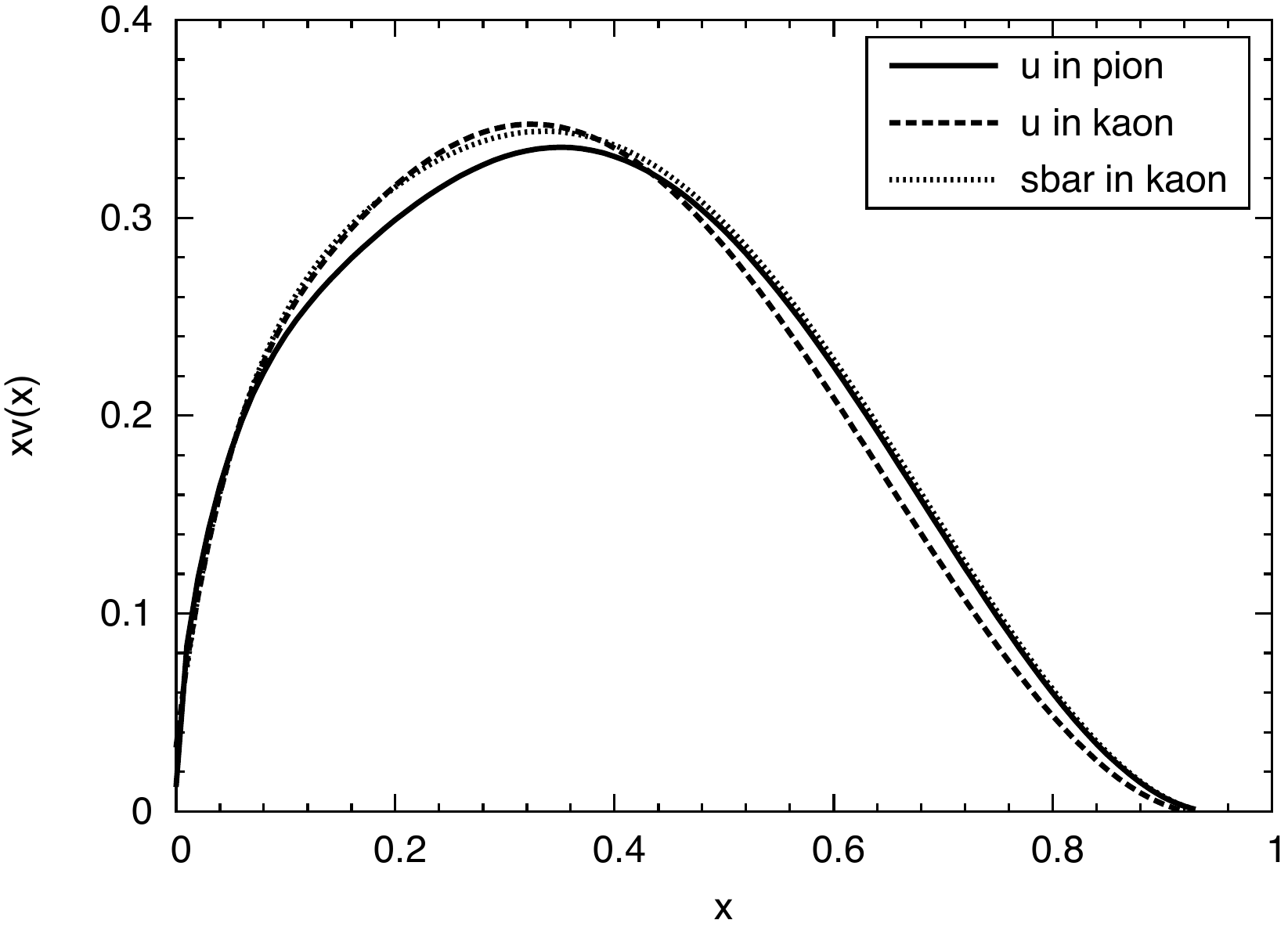}
\end{tabular}
\caption{
The dressed valence quark distributions as functions of Bjorken $x$.
Left:
The solid line shows the result taken from Ref.~\cite{Watanabe:2016lto}.
The dashed line represent the pion's distribution with some improved treatments (see the text for details).
The dotted and dashed-dotted lines denote the results for the $u$ and $\bar s$ quarks of the kaon, respectively.
These four results are obtained at  $Q^2=Q_0^2$.
Right:
The solid, dashed, and dotted lines depict the results for the pion's $u$ quark, the kaon's $u$ quark, and the kaon's $\bar s$ quark distributions at $Q^2 = 27$~GeV$^2$, respectively.
}
\label{fig:dress_total}
\end{figure}

Then, let us present our results for the dressed distributions at the initial and higher $Q^2$ in Fig.~\ref{fig:dress_total}.
%
%
First, we explain with the left panel in the figure the slight difference in the treatments of the pion's dressed distribution in this work.
In Ref.~\cite{Watanabe:2016lto}, the $\eta$ meson contribution was neglected, and the bare valence $u$ quark distribution of the kaon required for their numerical evaluations was simply replaced by the pion's.
Their results are shown by the solid line in the left panel.
On the other hand, in this work those two parts are improved, and the results are depicted by the dashed line.
As seen from the comparison, the difference between the two is extremely tiny, which justify the treatment in the previous work.
Nevertheless, the other calculations for the pion presented in this article are obtained with the updated one.
Focusing on the results for the kaon, our non-trivial finding is that the difference between $v^{K(u)}_{\rm dressed}$ and $v^{K(\bar{s})}_{\rm dressed}$ becomes slightly smaller compared with that from the bare ones demonstrated in Fig.~\ref{fig:bareVQDF}.
In other words, the meson cloud effects actually reduce the $SU(3)$ symmetry breaking effect in the distribution functions.
We also find that $v^{K(\bar{s})}_{\rm dressed}$ is almost identical to $v^{\pi}_{\rm dressed}$ when $x\ge 0.5$, although $v^{K(\bar{s})}_{\rm bare}$ is obviously larger than  $v^{\pi}_{\rm bare}$ in the region between $x=0.6$ and $0.8$.
After performing $Q^2$ evolution, the shapes of the curves change, but the qualitative relations among them remain the same.
Unlike the results presented in some preceding studies, e.g., the nonlocal chiral quark model~\cite{Nam:2012af,Nam:2012vm} or the Dyson-Schwinger equation~\cite{Chen:2016sno}, the observed size of the $SU(3)$ symmetry breaking in our analysis is clearly smaller.
To understand this, our further investigations on the contributions from each term in the expressions of the dressed distributions will be presented later in this section.
Before moving on to the part, we display the errors of the resulting distributions of the kaon, and present and discuss the moments calculated from the results.\\

\begin{figure}[tb]
\begin{center}
\includegraphics[width=0.55\textwidth]{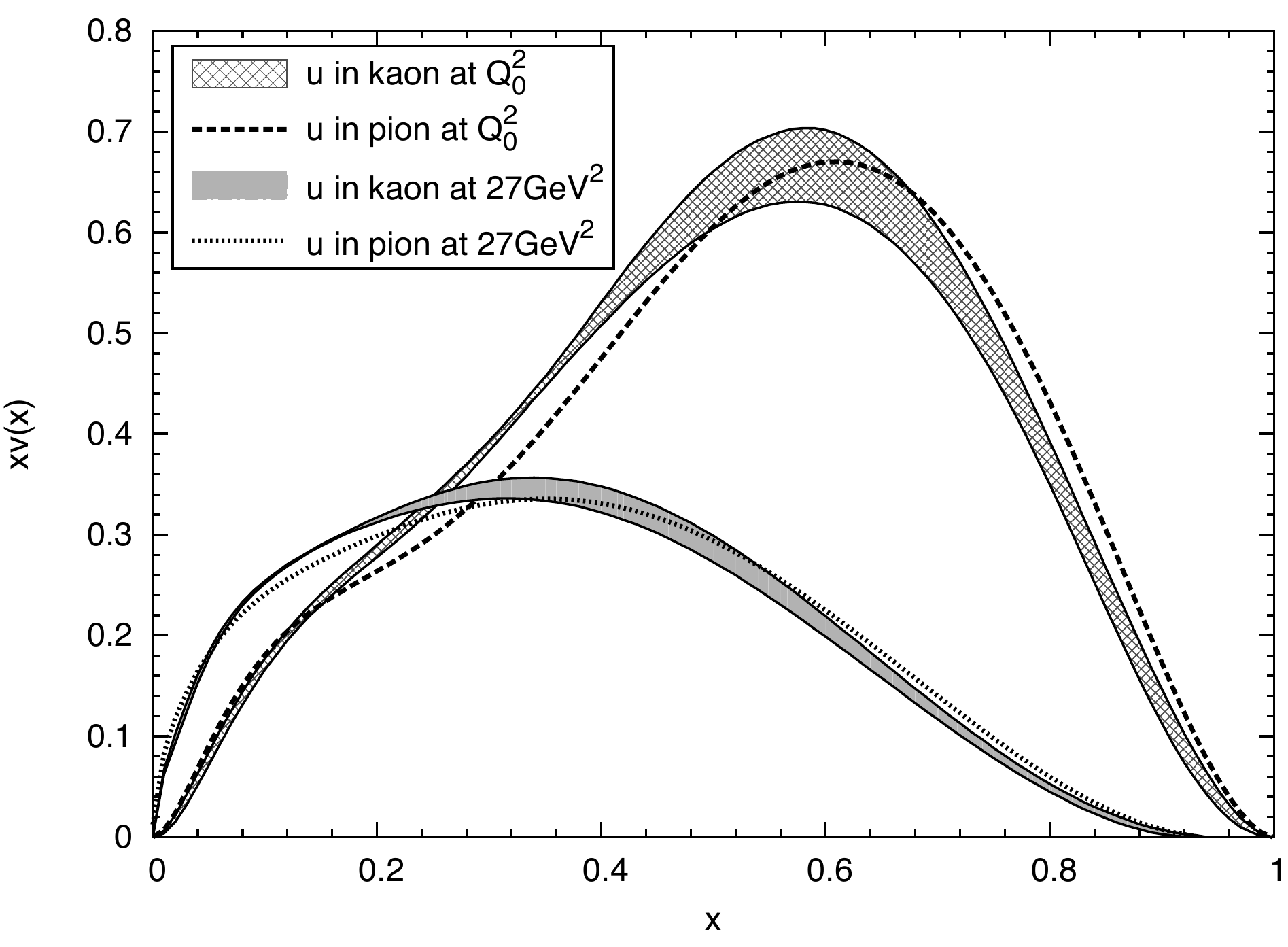}
\caption{
The dressed valence $u$ quark distributions of the pion and kaon as functions of Bjorken $x$.
The cross-hatched and shadowed bands show the 68\% C.L. $v^{K(u)}_{\rm dressed}(x)$ uncertainties at $Q^2 = Q_0^2$ and $27$~GeV$^2$, respectively.
The dashed and dotted curves denote the distributions for the pion at $Q^2 = Q_0^2$ and $27$~GeV$^2$, respectively.
}
\label{fig:dressed_uval_timesx_with_errors}
\end{center}
\end{figure}

\begin{figure}[tb]
\begin{center}
\includegraphics[width=0.55\textwidth]{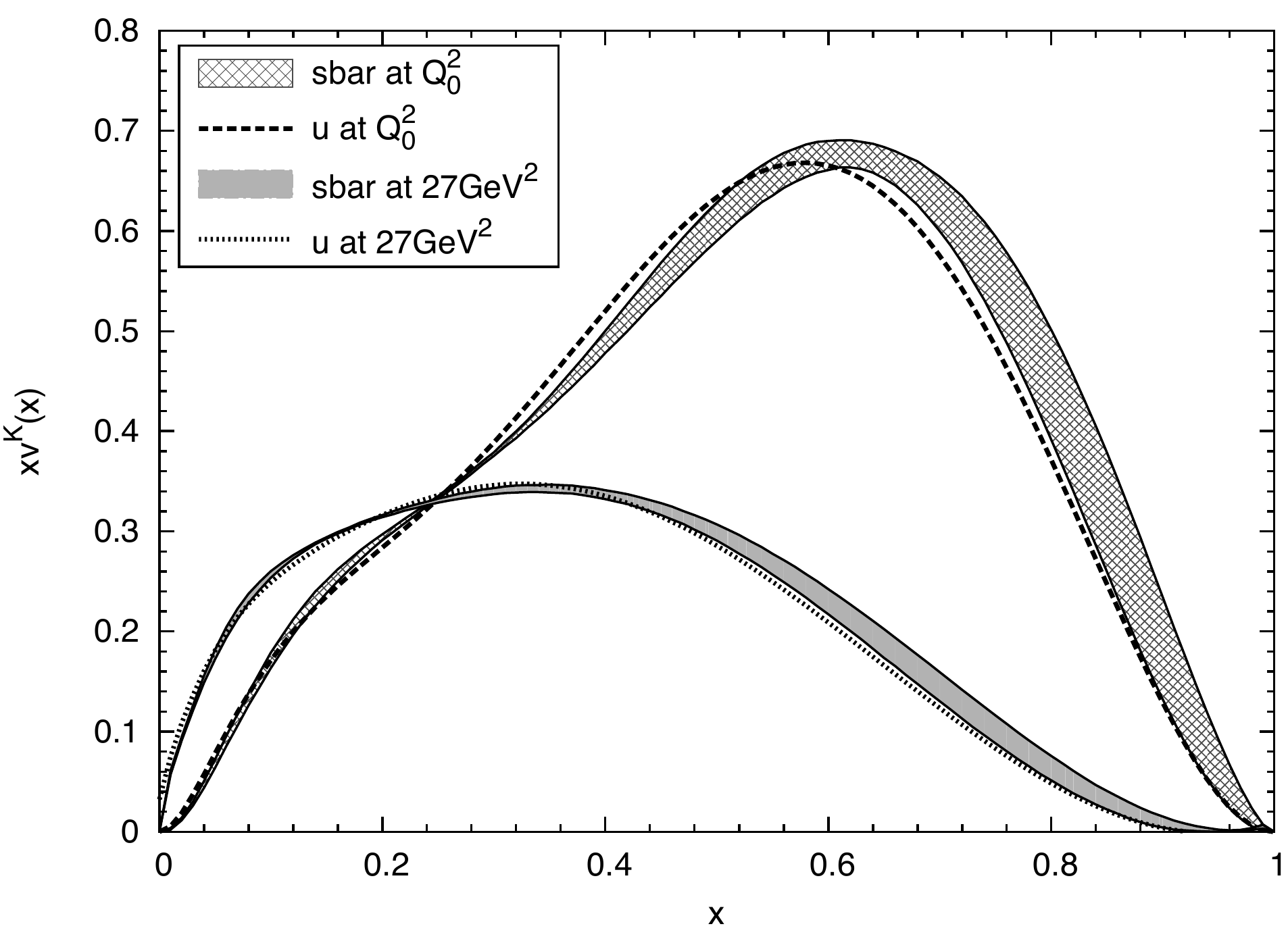}
\caption{
The dressed valence quark distributions of the kaon as functions of Bjorken $x$.
The cross-hatched and shadowed bands show the 68\% C.L. $v^{K(\bar{s})}_{\rm dressed}(x)$ uncertainties at $Q^2 = Q_0^2$ and $27$~GeV$^2$, respectively.
The dashed and dotted curves denote the $u$ quark distributions at $Q^2 = Q_0^2$ and $27$~GeV$^2$, respectively.
}
\label{fig:dressed_QDF_in_kaon_timesx_with_errors}
\end{center}
\end{figure}

First, we show in Fig.~\ref{fig:dressed_uval_timesx_with_errors}
%
%
the resulting dressed valence $u$ quark distribution of the kaon with its uncertainties, together with the pion's result for comparison.
One can see from the figure that $xv^{\pi}$ is smaller that $xv^{K(u)}$ in the region between $x = 0.2$ and $0.5$, but larger when $x \ge 0.7$ at the initial scale.
At $Q^2 = 27$~GeV$^2$, it is harder to find the difference between the two results, and they are almost identical to each other within the uncertainties particularly in the region where $x \ge 0.3$.
Next, we display the dressed valence quark distributions of the kaon with the uncertainties of the $\bar s$ quark distributions in Fig.~\ref{fig:dressed_QDF_in_kaon_timesx_with_errors}.
%
%
Although the differences are quite tiny, but one can find that $xv^{K(u)}$ is slightly larger than $xv^{K(\bar{s})}$ in the region between $x = 0.3$ and $0.5$, and smaller when the value of $x$ is in the range of $0.6 \sim 0.8$ at the initial scale.
We find that the two curves are almost identical to each other within the uncertainties in the whole $x$ region at $Q^2 = 27$~GeV$^2$.\\

Since it is useful to consider the moments of the obtained distribution functions to compare the results with those in the preceding studies, we present our resulting values in Table~\ref{table:moments}.
\begin{table}[tb]
\caption{
The first three moments obtained from the resulting dressed valence quark distributions at $Q^2 = 27$~GeV$^2$, compared with the results in Ref.~\cite{Chen:2016sno}.
}
\begin{center}
\begin{tabular}{| c | c || c | c | c |} \hline
$q$ & & $\langle x \rangle _q$ & $\langle x^2 \rangle _q$ & $\langle x^3 \rangle _q$ \\
\hline \hline
$v^\pi$ & this work & 0.23 & 0.094 & 0.048 \\
& \cite{Chen:2016sno} & 0.26 & 0.11 & 0.052 \\
\hline
$v^{K(u)}$ & this work & 0.23 & 0.091 & 0.045 \\
& \cite{Chen:2016sno} & 0.28 & 0.11 & 0.048 \\
\hline
$v^{K(\bar{s})}$ & this work & 0.24 & 0.096 & 0.049 \\
& \cite{Chen:2016sno} & 0.36 & 0.17 & 0.092 \\
\hline
\end{tabular}
\end{center}
\label{table:moments}
\end{table}
For comparison, the results of the Dyson-Schwinger equation based study~\cite{Chen:2016sno} are also shown in the table.
As to the results for $v^{\pi}$ and $v^{K(u)}$, our values are slightly smaller than theirs, but the differences are not large.
However, one can see the substantial difference between the two in the results for $v^{K(\bar{s})}$, which reflects the difference in the size of the $SU(3)$ flavor symmetry breaking observed in the analysis.\\

\begin{figure}[tb]
\begin{tabular}{cc}
\includegraphics[width=0.5\textwidth]{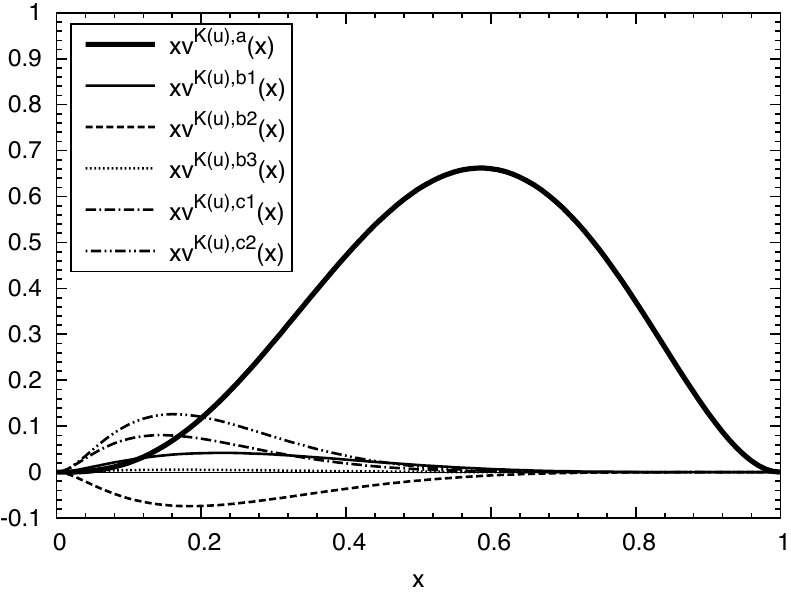}
\includegraphics[width=0.5\textwidth]{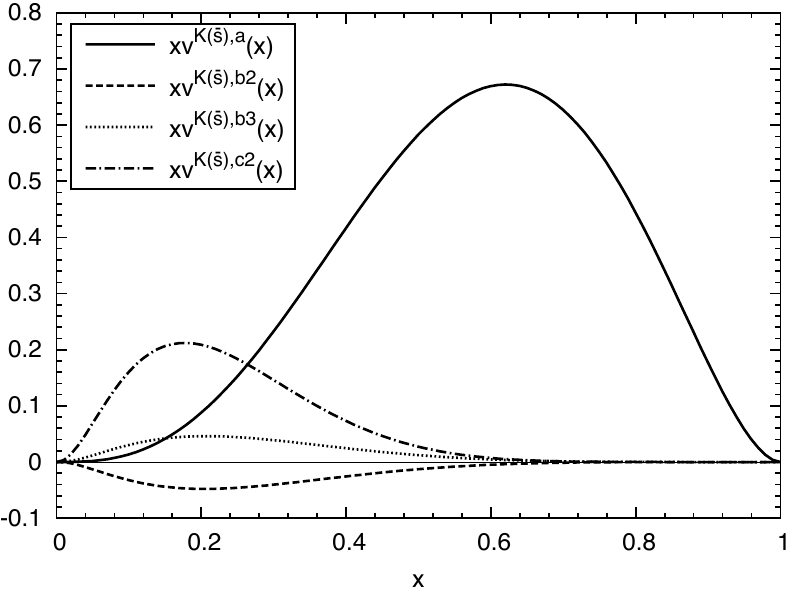}
\end{tabular}
\caption{
The contributions of each term in Eq.~\eqref{eq:eachterms_u} to $xv^{K(u)}_{\rm dressed}$ (left panel) and those from the terms in Eq.~\eqref{eq:eachterms_s} to $xv^{K(\bar{s})}_{\rm dressed}$ (right panel) at $Q^2=Q_0^2$ as functions of Bjorken $x$.
}
\label{fig:eachterms}
\end{figure}

The next step of our analysis is to make a detailed anatomy of the meson cloud effects on the valence quark distribution functions of the kaon at the initial scale.
The contributions from each term in Eqs.~\eqref{eq:eachterms_u} and \eqref{eq:eachterms_s} to dressed distributions are explicitly displayed in Fig.~\ref{fig:eachterms}.
%
%
The left panel is for $v^{K(u)}_{\rm dressed}$ and the right panel is for $v^{K(\bar{s})}_{\rm dressed}$.
Let us look at the $u$ quark case first.
The thick solid curve corresponding to $xv^{K(u),a}$ is just the bare valence quark distribution function multiplied by the renormalization constant $Z_u$.
It is obvious that this term gives the dominant contribution in the region where $x \ge 0.6$.
Thus our $u$ quark distribution behaves as $(1-x)^{\gamma}$ when $x\rightarrow 1$.
The contributions from the diagrams in Fig.~\ref{fig:diagrams}~(b) and (c) are significantly nonzero only in the region of $x\le  0.6$.
Among them, only $xv^{K(u),b2}$ is negative and the other terms are all positive.
One finds that $xv^{K(u),c2}$ is larger than $xv^{K(u),c1}$ in the whole region, and $xv^{K(u),c1}$ is larger than $xv^{K(u),b1}$ when $x \le 0.3$.
The $\eta$ meson contribution, $xv^{K(u),b3}$, is extremely small.
Then, let us look at the $\bar s$ quark case.
The role of $xv^{K(\bar{s}),a}$ is similar to that of $xv^{K(u),a}$, and this term gives the dominant contribution when $x \ge 0.6$.
$xv^{K(\bar{s}),b2}$ is negative, and $xv^{K(\bar{s}),b3}$ and $xv^{K(\bar{s}),c2}$ are positive in the whole region.
However, the magnitude of $xv^{K(\bar{s}),c2}$ is larger than double of that of $xv^{K(\bar{s}),b2}$.
Since the size of $xv^{K(\bar{s}),b3}$ is close to that of $xv^{K(\bar{s}),b2}$, the main contribution in the small $x$ region is from $xv^{K(\bar{s}),c2}$. \\

Now we can discuss the reason for the observed small size $SU(3)$ flavor symmetry breaking in our analysis.
In the large $x$ region, the difference between $v^{K(u)}_{\rm dressed}$ and $v^{K(\bar{s})}_{\rm dressed}$ at $Q^2=Q_0^2$ is basically from the difference between $Z_{u}v^{K(u)}_{\rm bare}$ and $Z_{s}v^{K(\bar{s})}_{\rm bare}$.
We find that $v^{K(\bar{s})}_{\rm bare}$ is larger than $v^{K(u)}_{\rm bare}$ in the region, and $Z_{s}$ is about 6\% smaller than $Z_{u}$.
Therefore, magnitudes of $v^{K(u)}_{\rm dressed}$ and $v^{K(\bar{s})}_{\rm dressed}$ become close to each other.
In other words, at large $x$ the dressing correction is negative, and the valence $\bar{s}$ quark receives a larger suppression compared with the $u$ quark.
Thus the difference between the resulting distributions becomes smaller.
In the small $x$ region, the difference is mainly from the different contributions of the dressing corrections shown in Fig.~\ref{fig:eachterms}.
Here the total size of the corrections is positive, and the valence $\bar{s}$ quark receives a larger enhancement mostly due to the contribution of $xv^{K(\bar{s}),c2}$.
Since $v^{K(\bar{s})}_{\rm bare}<v^{K(u)}_{\rm bare}$ in the small $x$ region, the resulting size of the $SU(3)$ flavor symmetry breaking becomes smaller compared with that we can see in the bare distributions. \\

\begin{figure}[tb]
\begin{tabular}{cc}
\includegraphics[width=0.5\textwidth]{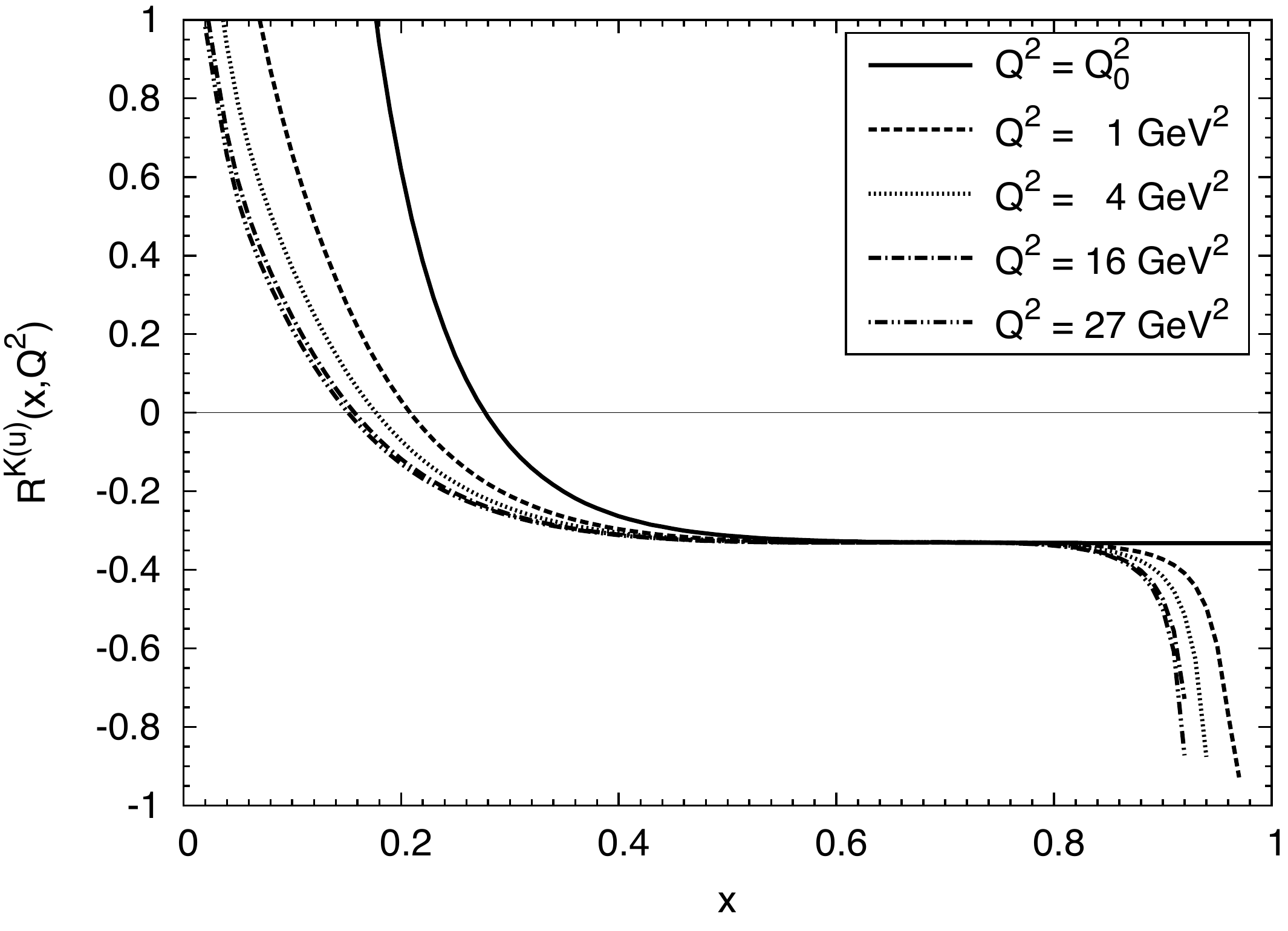}
\includegraphics[width=0.5\textwidth]{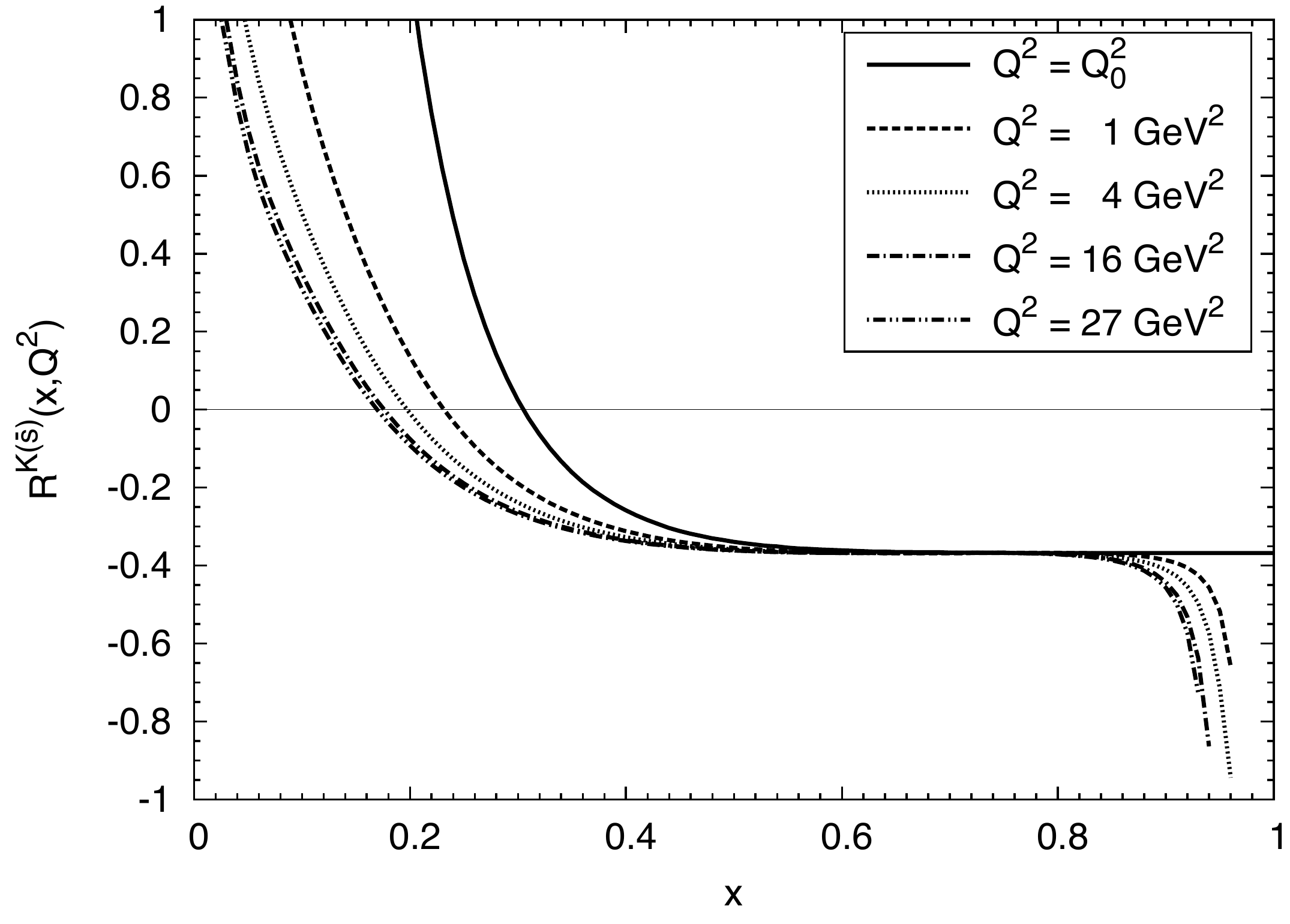}
\end{tabular}
\caption{
The ratios of the valence quark distribution functions, $R^{K(u)}(x,Q^2)$ (left panel) and $R^{K(\bar{s})}(x,Q^2)$ (right panel) defined in Eqs.~\eqref{eq:R_for_u} and \eqref{eq:R}, as functions of Bjorken $x$ for various $Q^2$.
The solid, dashed, dotted, dashed-dotted, and dashed-double dotted curves depict the $Q^2=Q_0^2$, 1, 4, 16, 27~GeV$^2$ cases, respectively.
}
\label{fig:correction_strength_from_original}
\end{figure}

Besides the above analysis, we also investigate the relative magnitudes of the dressing corrections to the constituent quarks of the kaon.
To do so, we define the following quantities,
\begin{align}
R^{K(u)} \left( {x,{Q^2}} \right) \equiv \frac{{v_{{\rm{dressed}}}^{K(u)} \left( {x,{Q^2}} \right) - v_{{\rm{bare}}}^{K(u)} \left( {x,{Q^2}} \right)}}{{v_{{\rm{bare}}}^{K(u)} \left( {x,{Q^2}} \right)}}, \label{eq:R_for_u} \\
R^{K(\bar{s})} \left( {x,{Q^2}} \right) \equiv \frac{{v_{{\rm{dressed}}}^{K(\bar{s})} \left( {x,{Q^2}} \right) - v_{{\rm{bare}}}^{K(\bar{s})} \left( {x,{Q^2}} \right)}}{{v_{{\rm{bare}}}^{K(\bar{s})} \left( {x,{Q^2}} \right)}}.
\label{eq:R}
\end{align}
Our results are displayed in Fig.~\ref{fig:correction_strength_from_original}.
%
%
The left panel is for $R^{K(u)}$ and the right panel is
for $R^{K(\bar{s})}$.
In the former one, the correction is simply the overall 33\% reduction due to the renormalization constant $Z_{u}$ in the large $x$ region at $Q^2=Q_0^2$, but at higher $Q^2$, $R^{K(u)}$ appears to be suppressed in the region of $x\ge 0.8$.
Furthermore, this suppression becomes more pronounced as $Q^2$ increases.
Usually the large $x$ behavior of the valence quark distribution function is characterized by some exponent $\delta$ such that $v(x)\sim (1-x)^{\delta}$ as $x\sim 1$.
The dressing effect makes the value of $\delta$ to be substantially larger as $Q^2$ increases.
The situation in the small $x$ region is rather different, and the dressing effect enhances the bare distribution significantly when $x\le 0.4$. However, the magnitude of this enhancement decreases as $Q^2$ increases.
Besides, there is a plateau when $x$ is in the range of $0.4 \sim 0.8$ where the dressing effect is just the overall 33\% suppression for any value of $Q^2$.
This general feature can also be seen in $R^{K(\bar{s})}$, while the size of the overall reduction becomes about 37\% and the enhancement in the small $x$ region is slightly larger compared with the $u$ quark case. \\

Finally, we make a remark on the parameter dependencies of the renormalization constants which determine the size of the overall suppression and affect the determination of the bare distribution functions.
As mentioned earlier, we utilized a standard set of the model parameters with which one can reproduce the empirical value of the Gottfried sum rule.
If we increase the constituent quark masses or the cutoff parameter $\Lambda$, the sum rule value decreases.
For instance, if we increase the masses by 10$\%$ and choose an appropriate value for $\Lambda$, we can still obtain the same sum rule value.
In this case, the obtained values of the renormalization constants change, but the differences are less than 2$\%$.
Hence, we conclude that the parameter dependencies are not crucial in this study.

\section{Conclusion}
In this article, the valence $u$ and $\bar{s}$ quark distribution functions of the $K^+$ meson have been studied, and the meson cloud effects on its constituent quarks have been discussed in detail in the framework of the chiral constituent quark model.
We judiciously chose the bare valence quark distributions to generate the dressed distributions which agrees with the phenomenologically satisfactory $v^{\pi}$ and the experimental data of the ratio $v^{K(u)} / v^{\pi (u)}$.\\

We found that the resulting distributions show an obviously small $SU(3)$ flavor symmetry breaking, compared with results of the preceding studies based on other approaches.
In our results, the three distributions, $v^{\pi}$, $v^{K(u)}$, and $v^{K(\bar{s})}$, are close to each other at both the initial scale and the higher $Q^2$.
The similar observation can be seen from the small differences among the moments calculated with them.
To find out the reasons for this, we then investigated the contributions of each term in the expressions of the dressed distribution functions.
The observed general features of the dressing corrections are consistent with the findings in the previously studied pion case, and there is no crucial difference in between the $v^{K(u)}$ and $v^{K(\bar{s})}$ cases.
The contributions from the renormalization constants are dominant in controlling the moments, and actually the difference between $Z_u$ and $Z_{s}$ is small, although the $SU(3)$ flavor symmetry breaking effects are clearly introduced in the model parameters such as the constituent quark masses.
This is our non-trivial finding in this study, and contributes in part to the small symmetry breaking.\\

The analysis presented in this article is model dependent, and making modifications, such as employing more complicated functional forms for the bare quark distributions, is possible.
However, it is also true that the currently available experimental data for the kaon have large uncertainties, which makes it quite difficult to impose constraints on the adjustable parameters in the model.
It should also be noted here that our results for the kaon depend on the distribution function of the pion.
Not a few studies on the pion PDFs have been done so far, but they have not yet been pinned down because of a lack of data.
However, the pion-induced Drell-Yan experiment is currently performed by COMPASS Collaboration at CERN~\cite{Gautheron:2010wva}.
The forthcoming results from this experiment hopefully will improve our understandings about the pion structure in the near future.
Furthermore, the high intensity kaon beam might be available at some experimental facilities such as COMPASS and J-PARC~\cite{Fujioka:2017gzp} in the future.
Via the kaon-induced Drell-Yan experiment at these facilities, our predictions presented here could be tested.

\section*{Acknowledgments}
We would like to thank Hsiang-nan Li, Hai-Yang Cheng, and Wen-Chen Chang for their valuable suggestions and comments.
This work was supported in part by the Ministry of Science and Technology of Taiwan
 (MOST 105-2811-M-033-004, MOST 105-2112-M-033-004, and NSC 102-2112-M-033-005-MY3)
 and the U.S. National Science Foundation (PHY-1505458).

\end{document}